# Forescattered electron imaging of nanoparticles in a scanning electron microscopy

Junliang Liu*, Sergio Lozano-Perez, Phani Karamched, Jennifer Holter, Angus J. Wilkinson, Chris R. M. Grovenor

Department of Materials, University of Oxford, Parks Road, OX1 3PH, United Kingdom

*Corresponding author: bryanchina@hotmail.com

**Abstract**

In this study, we have used a Zr-Nb alloy containing well-defined nano-precipitates as a model material in which to study imaging contrast inversions (atomic number or diffraction contrast) observed with the forescattered electron imaging system, ARGUS™, in a scanning electron microscope (SEM) when imaging a thin foil in a transmission geometry. The study is based on Monte Carlo simulations and analysis of micrographs experimentally acquired under different imaging conditions. Based on the results, imaging conditions that enhance atomic number or diffraction contrast have been proposed. Data acquired from the ARGUS™ imaging system in SEM has also been compared with results from standard transmission electron microscopy and scanning transmission electron microscopy imaging of the same material. These results demonstrate the capability of the ARGUS™ system to investigate microstructures in nano-scale materials.

## 1. Introduction

The size and number densities of nano-particles dispersed in solid metallic matrices are well-known to be one of the most important factors that control the physical properties of in a wide range of engineering materials. In the Zr alloys used as a nuclear cladding material, second phase particles (SPPs) of different size, number densities and chemical composition can directly affect their mechanical properties, irradiation response and corrosion performance [1]. For instance, in Zircaloy-4, Motta et al [2] have shown that increasing the size of SPPs may improve the corrosion resistance. In the Zr-Nb system, most of the SPPs are small β-Nb particles (diameter <150 nm) homogeneously distributed inside the grains [3], and these alloys have shown better corrosion resistance than Zircaloy-4 in which most of the SPPs are Zr-Cr-Fe type [4]. In service as a fuel cladding alloy, radiation effects like irradiation-induced elemental segregation, amorphization and dissolution can change the size and distribution of different kinds of SPPs, and further affect the in-reactor corrosion performance [5]. Precise determination of the size and number density of nano-particles in nuclear cladding materials before and after radiation damage is therefore crucial for the development of reliable mechanistic models for predicting the safe lifetime of Zr alloys in-reactor.

Advances in high-resolution characterisation techniques such as scanning electron microscopy (SEM), Energy-dispersive X-ray (EDX), transmission electron microscopy (TEM) and atom probe tomography (APT) have enabled the study of materials at the nanoscale [6-8]. Microstructures such as embedded nano-particles, phase interfaces and grain boundaries can be imaged using backscattered electrons (BSE) [9, 10] and secondary electrons (SE) [11] in an SEM with adequate signal-to-noise ratios (SNR). However, the large electron interaction volume in a normal bulk SEM specimen limits the spatial resolution, especially when using BSE for imaging. Although the use of SE images would greatly reduce the interaction volume, SEs provide more surface topographical contrast than Z (atomic number) contrast, making it difficult to distinguish nanoparticles from surface features or contamination. TEMs are capable of achieving a spatial resolution below 1 nm from embedded particles in electron transparent foils, but when the TEM images contain other features like dislocations, surface oxides, and diffraction-related bend contours, it can be difficult to identify the precise shape of embedded nano-particles. The dominance of Z-contrast in high annular dark-field (HAADF) imaging in a scanning transmission electron microscope (STEM) makes it ideal for the characterisation of embedded nano-particles with high spatial resolution, irrespective of the surface topography. However, the maintenance/operation of a STEM compared with a normal SEM is generally complex and costlier.

In this paper, we present some recommendations on an experimental strategy to achieve the optimum contrast inversions (atomic number or diffraction contrast) when using the forescattered electrons in an SEM for imaging nano-particles embedded in a thin foil. This work is based on Monte Carlo simulations combined with experimental observations of electron interactions with embedded Nb-rich nano-particles in a thin foil and the analysis of SEM, TEM and STEM images from a model Zr-Nb alloy.

## 2. Materials and methodologies

### 2.1. Materials

The Zr-1.0Nb alloy in sheet format was provided by Westinghouse after a final annealing step at 560 °C, which leads to recrystallized α-Zr grains with the Nb mostly in solid solution and small Nb-containing particles. These SPPs are homogeneously distributed inside the grains, and are mostly β-Nb particles with a few $Zr(Nb,Fe)_2$ Laves phase particles. Typical microstructures of Zr-Nb alloys can be seen in Fig. S1. Three of the $Zr(Nb,Fe)_2$ Laves particles of larger size are highlighted by dashed rectangles. The β-Nb particles are of smaller size with an average diameter of ~50 nm. A 3D representative model was then built based on the average size of β-Nb particles, and will be described in section 2. 3.

### 2.2. Experimental methods

Electron transparent thin foil specimens containing Nb particles for analysis by forescattered electron imaging were prepared using the in-situ FIB lift-out method on a Zeiss Crossbeam 540 FIB/SEM system. Following the steps described in [12], lamellar specimens were lifted out and welded to a 3 mm Cu grid using an in-situ micro-manipulator. Thinning with a gradually decreasing milling current of 1500-100 pA at 30 kV was performed on both sides of the foil, and final thinning to electron transparency and cleaning was performed at 5 kV and 200 pA. Conventional TEM images were acquired on these specimens using a JEOL 2100 microscope operating at 200 kV, and STEM images using a JEOL ARM 200F microscope operating at 200 kV with a pixel size of ~1 nm.

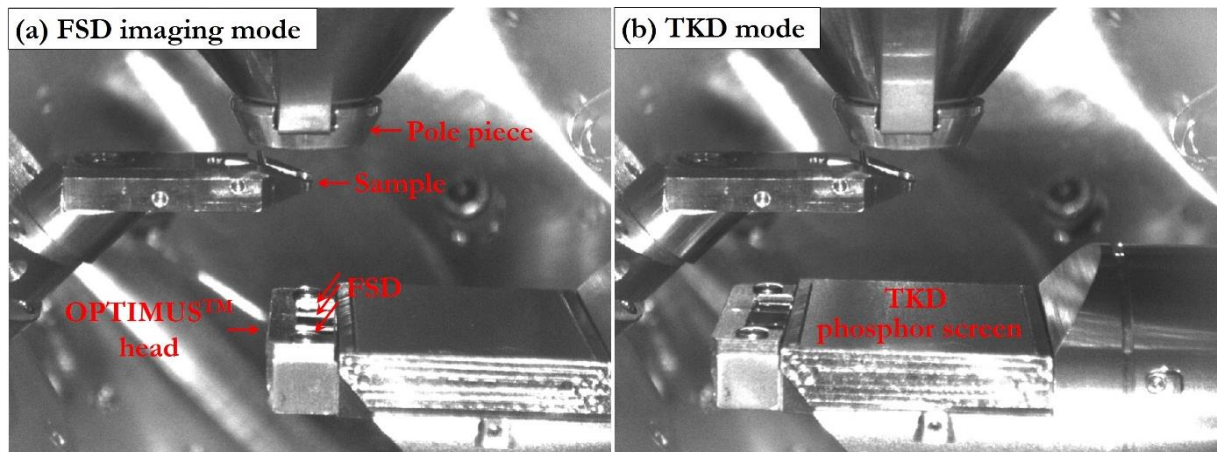

Fig. 1 In-chamber camera images of typical experimental configurations for (a) FSD imaging mode, (b) transmission Kikuchi diffraction mode

Forescattered electron imaging was carried on a Zeiss Merlin FEG-SEM system equipped with a Bruker e-Flash high-resolution EBSD system with an OPTIMUS™ head, Fig. 1. The OPTIMUS™ head consists of a phosphor screen and three forescattered electron detectors (FSD). When using the FSD imaging mode, Fig. 1(a), the center of low-angle forescattered electron detector (labelled as 'LAFSD') is aligned normal to the optic axis for generating bright-field images, and the two high-angle detectors (labelled as 'HAFSD') for dark-field images. When working at TKD mode, the phosphor screen is aligned normal to the optic axis for collecting the on-axis transmission Kikuchi diffraction patterns, Fig. 1(b). More information about orientation mapping at TKD mode can be found in [13-15]. Fig. 2 shows schematic three-dimensional-view of FSD detector geometries and an enlarged image of the configurations of OPTIMUS™ head. Different from angular dark-field (ADF) detectors used in STEM, the three FSDs in ARGUS™ system are rectangular. Even though slightly different collection semi-angles are associated with x and y directions, we believe this will not induce significant changes in the image contrast and thus the calculation of corresponding collection semi-angles at different detector distances ($d_d$) is only carried along the x-direction, Fig. 2 (a and c), based on the following equations,

$$\beta_1 = \arctan\left(\frac{d_1}{d_d}\right)$$

$$\beta_2 = \arctan\left(\frac{d_3}{d_d}\right) - \arctan\left(\frac{d_2}{d_d}\right)$$

where $\beta_1$ and $\beta_2$ are the semi-angles of the LAFSD and HAFSD respectively, $d_1$=2.6 mm, $d_2$= 4.3 mm, $d_3$= 9.5 mm. Result are summarized in Table 1.

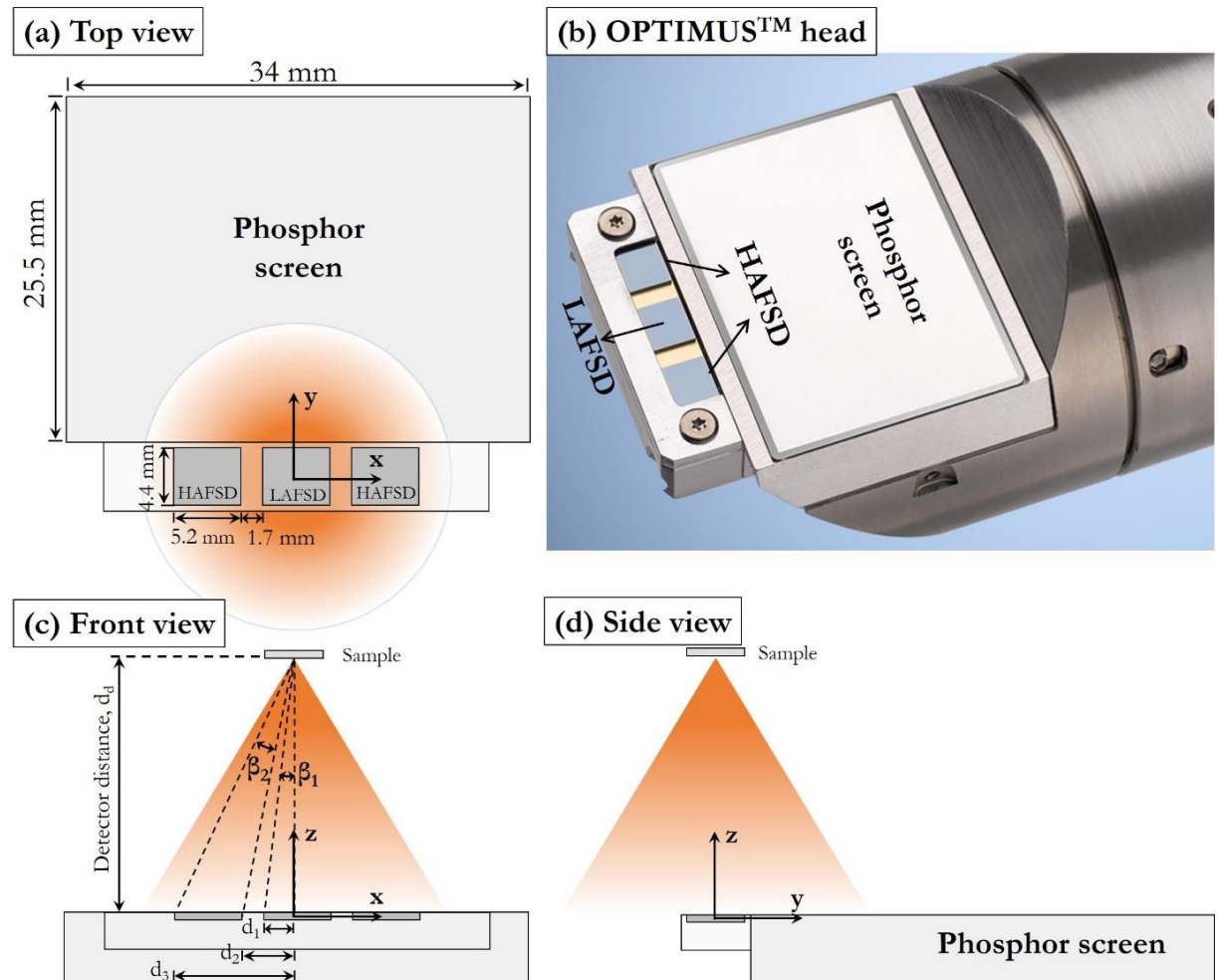

Fig. 2 (a, c and d) Schematic three-view of the OPTIMUS™ head working at FSD imaging mode with key parameters, detector size, working distance ($d_W$), detector distance ($d_d$) and collection angles ($\beta_1$ and $\beta_2$). The intensity distribution of transmitted electrons is shown schematically by the color gradients. (b) an enlarged image showing the configurations of OPTIMUS™ head [16].

Table 1 FSD collection semi-angles (mrad) as a function of detector distances

| $d_d$ (mm) | 5 | 10 | 15 | 20 | 25 | 30 | 35 |
|---|---|---|---|---|---|---|---|
| LAFSD | 0-480 | 0-250 | 0-170 | 0-130 | 0-100 | 0-90 | 0-70 |
| HAFSD | 710-1100 | 410-760 | 280-570 | 210-440 | 170-360 | 140-310 | 120-270 |

### 2.3. The methodology of Monte Carlo Simulations

Monte Carlo simulations, using random numbers and probability distributions to represent the physical interactions between the electron and the sample, are widely used to understand the capabilities of electron microscopes [17-19]. Casino Monte Carlo software v3.3 [19] which describes electron trajectories by discrete elastic scattering events, and inelastic events approximated by a mean energy loss model between two elastic scattering events, was used in this study to simulate forescattered electron imaging. Using the electron transport three-dimensional feature of Casino v3.3, the beam and scanning parameters allow the simulation of realistic images [19]. Sample 3D model was built based on the size of the β-Nb nano-particles shown in Fig. S1. A niobium sphere ($\varrho$=8.56 g.cm$^{-3}$, plasmon energy = 9.12 eV and work-function =4.3 eV) with a diameter d of 50 nm is introduced at mid-thickness of a Zr foil ($\varrho$=6.49 g.cm$^{-3}$, plasmon energy = 16.6 eV and work-function =4.05 eV) with a thickness of 100 nm, as shown in Fig. 3. The Nb particle vary only 1 in atomic number compared to the matrix (Nb Z= 41, Zr Z= 40), though have significantly higher density. Pre-calculated values of the electron elastic and inelastic collision cross sections and the mean energy loss model in the software were used for the calculations. Each simulated image contains 30 x 30 pixels, with a pixel size of 4 nm. A beam diameter of 3 nm was selected which is approximately equal to that used experimentally and, on average for each point in the calculated images 2 x 10$^5$ electron interactions were simulated. Accelerating voltages of 10 kV, 20 kV and 30 kV have been used. The shot noise of the electron gun was included by varying the nominal number of electrons per point based on the noise characteristics of a field emission gun [19].

### 3. Results and discussion

#### 3.1. Contrast inversions revealed by Monte Carlo simulations

The cross-sectional view of electron trajectories in Fig. 3 shows an example of interactions between the electron beam ($E_0$ = 30 kV) and a niobium nanoparticle embedded in a Zr matrix. As expected, the electron beam is clearly seen to broaden after interacting with the sample, and the degree of electron scattering depends mostly on material composition, accelerating voltages and sample thickness. By collecting the electrons forescattered over different angular ranges, the modelled image can exhibit different degrees of Z contrast.

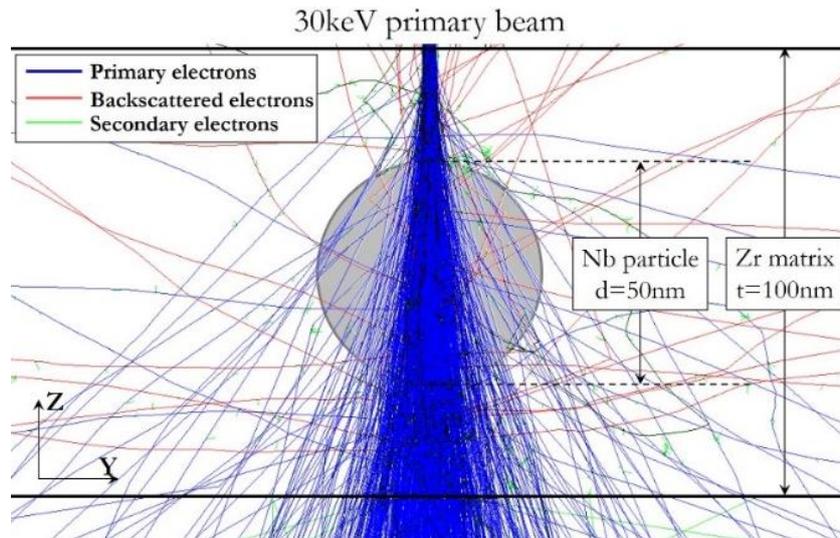

Fig. 3 Cross-sectional view of the 3D sample model and electron trajectories simulated by the Monte Carlo method

Above a certain high angle we expect the dominant signal to be from Rutherford scattering and the intensity of each pixel in the image will to be proportional to $tZ^2$, where t is the foil thickness and Z is the average atomic number [20]. The modelled angular distribution of forescattered electrons as a function of the incident electron energy is plotted in Fig. 4(a). As expected, more electrons are scattered to higher angles when the accelerating voltage is decreased from 30 kV to 10 kV, and as a result the beam broadening effects are more obvious for primary beams of lower energies. By using a detector distance/accelerating voltage combination that gives large collection angles (e.g. β > 400 mrad when $E_0$ =30 kV) most of the collected electrons will be Rutherford scattered [20]. If the collection angles are large enough, signals from coherent elastic scattering (diffraction) can also be much reduced, or even eliminated, and thus we can generate images showing Z-contrast exclusively with Rutherford scattered electrons [20]. Fig. 4(b) shows a series of images (30 kV accelerating voltage) simulated by collecting different signals, SE, BSE and electrons forescattered over different angular ranges. In each case the image intensity is normalized by the number of incident electrons simulated. The contrast C was calculated according to $C = \frac{I_P - I_M}{I_M}$ [21], where $I_M$ and $I_P$ are the average intensities of pixels in the particle and matrix, respectively. The SE image provides less compositional contrast [22] making it difficult to distinguish the nanoparticle, Fig. 4(b-1). BSE imaging, Fig. 4(b-2), allows structures with different compositions to be distinguished, thus a stronger contrast 0.145, but the intensity is relatively low, as more than 90% of the electron beam is transmitted. This decrease in signal intensity increases the effect of noise on the image resolution. When the electron beam passes through the heavier Nb (Z=41) particle, more electrons will be scattered to higher angles compared with in the Zr (Z=40) matrix. As a result, by collecting the high-angle scattered electrons (e.g. β > 400 mrad), the image shows stronger Z-contrast, Fig. 4(b-4), compared with collecting the low-angles scattered electrons (e.g. β < 100 mrad), Fig. 4(b-3). Signals from coherent elastic scattering (i.e. diffraction contrast) are not taken into consideration in these simulations [19].

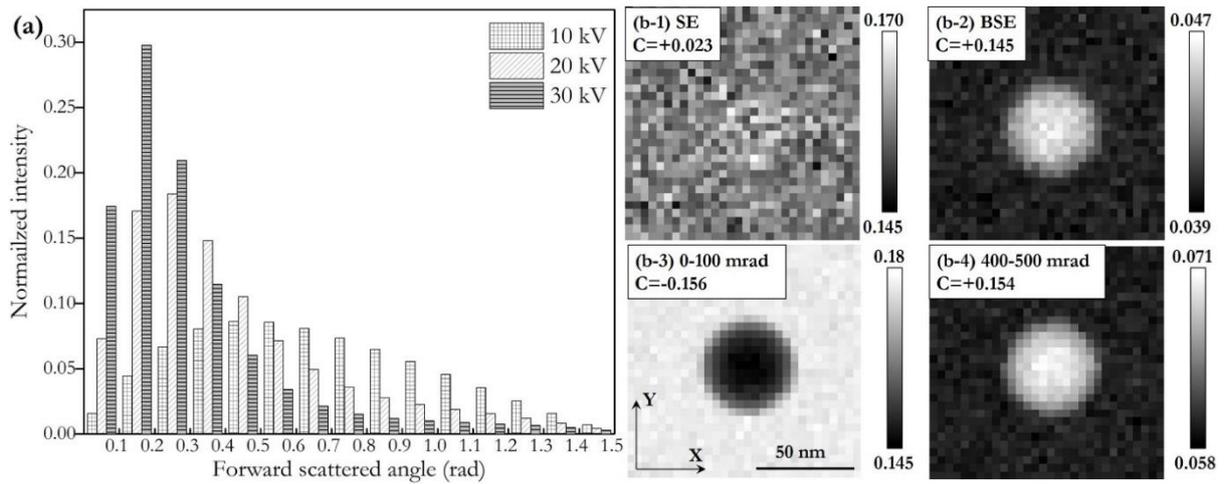

Fig. 4 Monte Carlo simulations showing (a) angular distributions of forescattered electrons and (b) images formed by 30 kV accelerating voltage and by collecting secondary electrons (SE), backscattered electrons (BSE) and forescattered electrons within 0-100 mrad and 400-500 mrad respectively.

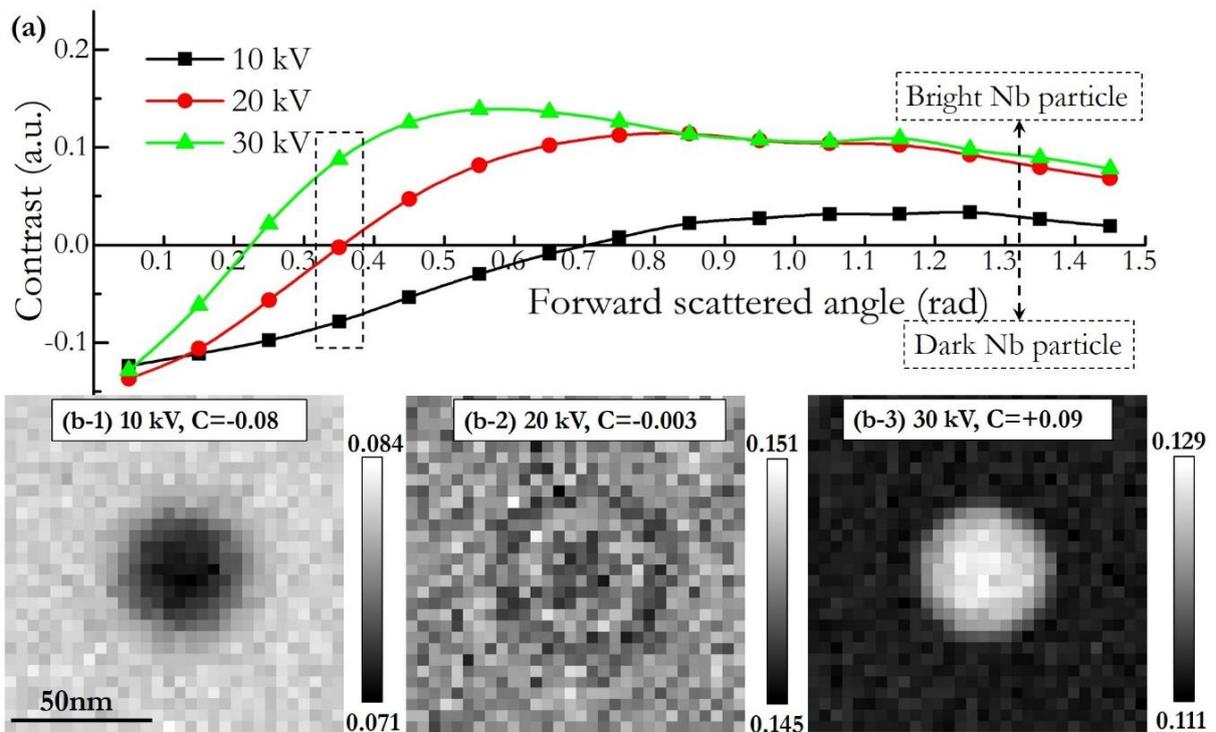

Fig. 5 (a) Monte Carlo simulations showing contrast inversions as a function of collection angles and accelerating voltages. (b) examples of simulated images using electrons scattered in the range 300-400 mrad.

Fig. 5 shows the contrast changes in images simulated by collecting forescattered electrons from different scattering angles and using different accelerating voltages. Contrast reversals can occur depending on the collection angles, Fig. 5(a). The critical values for contrast reversals increase with decreasing accelerating voltages, and as a result images acquired at different accelerating voltages can show different contrast (e.g. Z-contrast or diffraction contrast) even using the same detector distance (corresponding to specific collection angles). Examples of simulated images from the same collection angles (300-400 mrad) but using different accelerating voltages are shown in Fig. 5(b). For each image, the brightness range was equalized, and the intensity of each pixel normalized by the number of incident electrons simulated. The particle is visible in Fig. 5(b-1) and (b-3) but with inverse contrast. In comparison, for this specific range of collection angles, the particle is almost invisible at 20 kV, Fig. 5(b-2). The diode detectors used in realistic experiments generally have a cut-off threshold energy, and a linear scaling with electron energy above this [23]. The energy distribution of the transmitted electrons simulated in this works is shown in Fig. S2. The energy variance of forescattered electrons are very small especially for 20 and 30 keV incident electrons. For example more than 90% of the forescattered electrons for 30 keV primary electrons have their energy ranging from 29.5 keV to 29.6 keV. We believe this small energy variance will not induce significant changes in the image contrast, and will not go too far in this direction.

We have calculated and summarized the collection angles of the LAFSD and HAFSD in Table 1. By modelling the collection of electrons from those angles we are able to simulate more realistic images to compare with our experimental results. The contrast of LAFSD and HAFSD imaging as a function of detector distance and accelerating voltages is plotted in Fig. 6(a). In the range of available detector distances, LAFSD imaging always shows inverse Z-contrast regardless of the accelerating voltages. If the crystalline nature a typical sample is considered, LAFSD imaging could also show strong diffraction contrast [24], which has the potential to significantly alter the contrast from that simulated. In the case of HAFSD imaging, the degree of Z-contrast depends on the selected combination of accelerating voltages and detector distances. The critical detector distances($d_d^*$) below which Z-contrast shows the Nb particle as bright is forward to increase with increasing accelerating voltage. In other words, when using low accelerating voltages (e.g. 10 kV), the smaller the detector distance the more Z-contrast we have in the images. Using the threshold values obtained from Fig. 6(a), an expression for the critical detector distances to image Nb nanoparticles in a Zr matrix has been derived by polynomial regression, Fig. 6(b). The mathematical form of the regression function has no physical meaning, but could offer guidelines for setting specific detector distances when use the FSD system to image nanoparticles at different acceleration voltages.

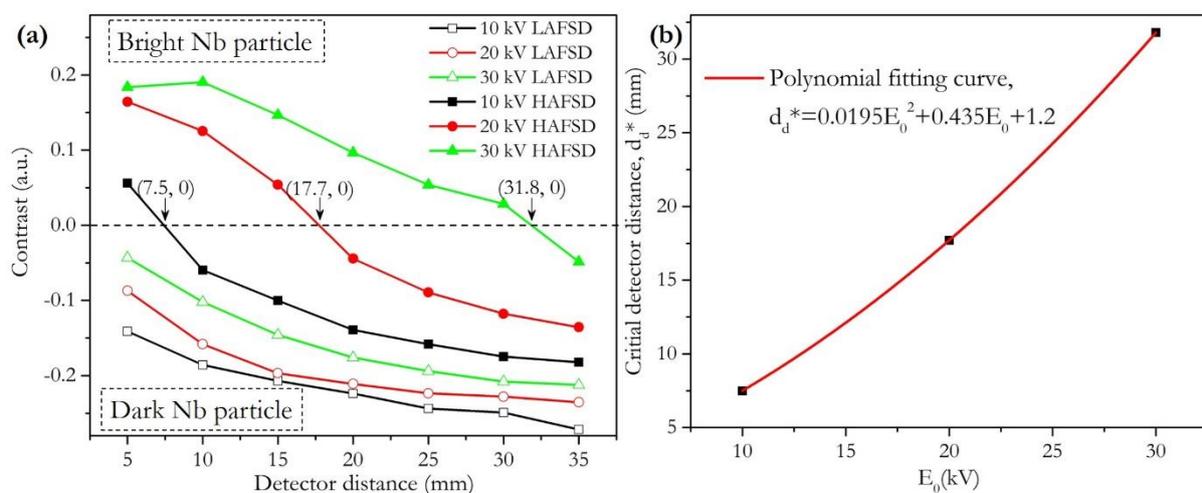

Fig. 6 (a) Monte Carlo simulations showing the imaging contrast as a function of FSD detector distances and accelerating voltages, and (b) polynomial fitting curve of critical detector distance ($d_d^*$) as a function of accelerating voltages to show Z-contrast when using the HAFSD to image the Nb-particle embedded in a Zr matrix

### 3.2. Experimental results

Fig. 7 shows a series of experimental images acquired from the same region in a Zr-Nb sample by HAFSD and LAFSD imaging at different detector distances and acceleration voltages. The second phase particles show a higher contrast in the Z-contrast images acquired at a detector distance of 10 mm for both 20 kV and 30kV accelerating voltages. At 20 mm distance, the Z-contrast becomes less visible in the HAFSD image generated with an accelerating voltage of 20 kV, and with increasing detector distances the image start to show diffraction contrast. In summary, when Z-contrast images are required, the smaller the detector distance used the higher the Z-contrast can be achieved. A similar tendency is also seen in the images generated with an accelerating voltage of 30 kV, although the contrast reversal occurs at a larger detector distance, ~30 mm, in good agreement with the Monte Carlo simulations. The relative contrast of a particle imaged at different detector distances and accelerating voltages was measured and plotted in Fig. 7(b). Unfortunately the electronic signal corresponding to no electrons reaching the detector was not measured so absolute contrast measurements cannot be constructed. However, the tendency of the relative contrast evolution with detector distances and accelerating voltages is seen in good agreement with that expected from the simulations in Fig. 6(a). The absolute values of contrast revealed by the simulations, Fig. 6(a), is around a factor of 6 lower than the relative contrast experimentally measured, Fig. 7(b). This is somewhat expected since the simulation account for a true zero signal level while the brightness and contrast settings used for the experimental acquisitions prevent this.

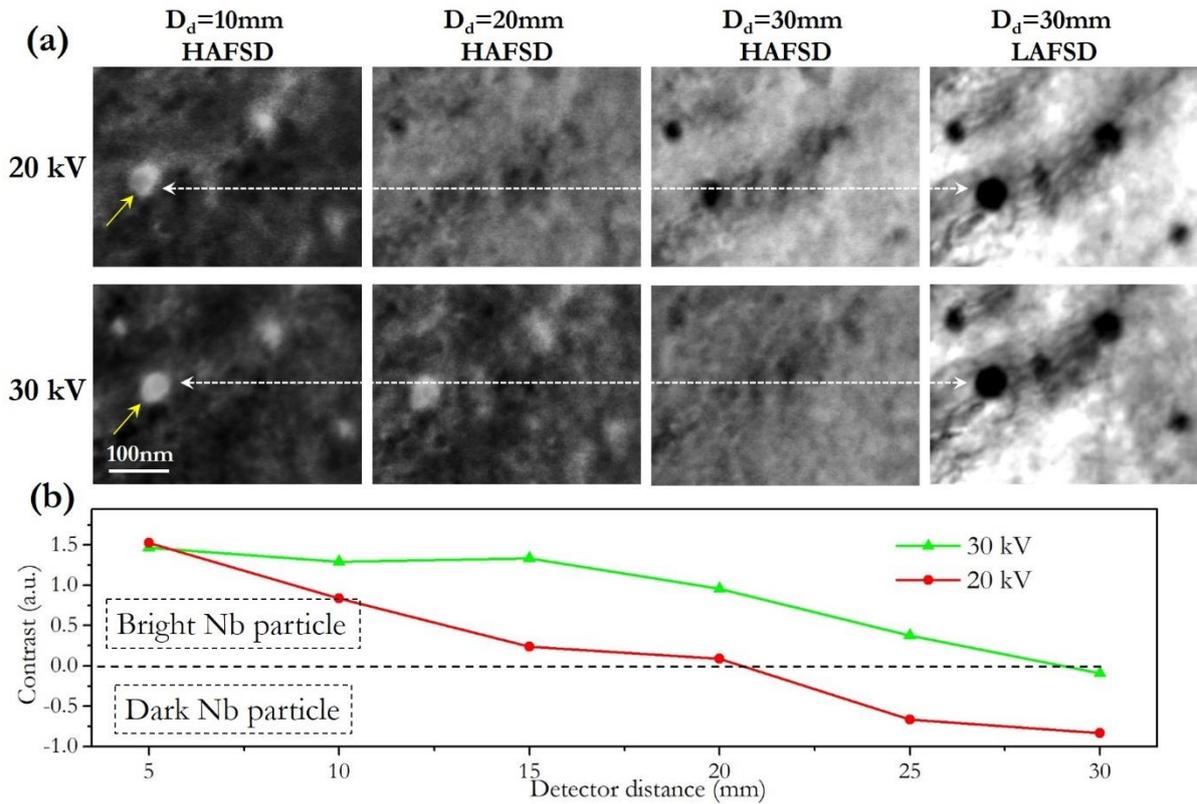

Fig. 7 (a) HAFSD and LAFSD images of the same region in a Zr-Nb sample using different detector distances and acceleration voltages, (b) the measured contrast of the same particle highlighted by yellow arrows in (a) at different detector distances

The contrast evolution can also be reflected by the distribution of forescattered electrons which carry the diffraction information. A transmission Kikuchi pattern (TKP) was acquired from this same specimen at 30 keV, Fig. 8. It can be seen that the majority of diffraction signal in the TKP is within a radius ~12 mm from the central direct beam, Fig. 8(a). Taking consider the detector distance used ($d_d$=20 mm) for acquiring this TKP, it can thus be estimated that forescattered electrons carrying diffraction information (or contribute to the diffraction contrast in an image) mainly distributed in a cone with a semi-angle ~540 mrad as shown in Fig. 8(b). When FSD imaging is carried at larger detector distances, HAFSD will have larger possibilities to collect diffracted electrons and show weak Z contrast. By contrast, when smaller detector distance is used, e.g. $d_d$< 10 mm, limited diffracted electrons will be collected by HAFSD and the images will thus show strong Z contrast.

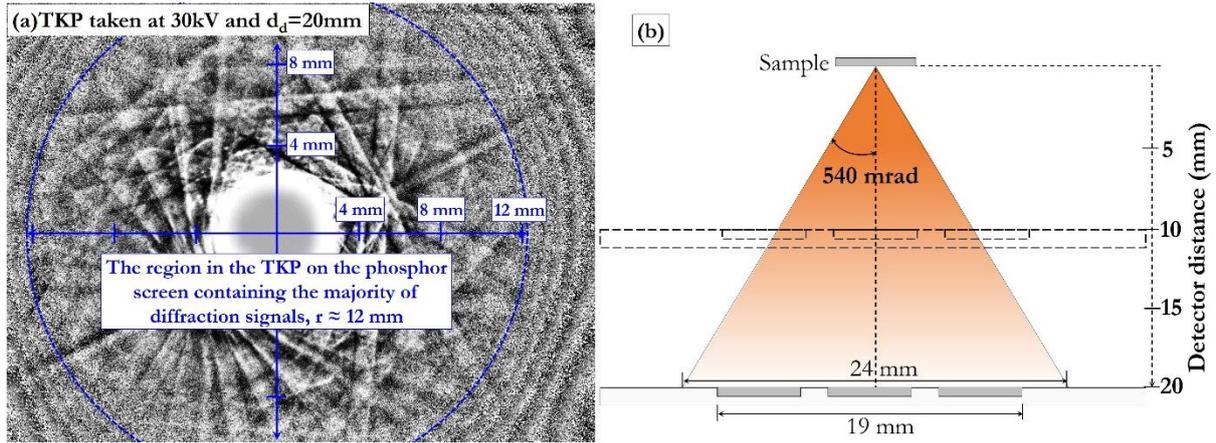

Fig. 8 (a) A transmission Kikuchi pattern acquired at a detector distance of 20 mm and 30 keV (the region containing strong diffraction signals is highlighted by the blue circle with the scales marking its size) and (b) a schematic showing the angular range of the forescattered electrons that carrying diffraction information and the intensity of these electrons is shown schematically by the color gradients, for example, at a detector distance of 20 mm, the range of diffracted signal is distributed within a radius ~12 mm from the central direct beam in the TKD phosphor screen.

Comparisons of the contrast shown by embedded Nb nanoparticles in the Zr-Nb alloy imaged by the FSD imaging system and the other techniques are shown in Fig. 9. As a result of the large difference in the accelerating voltages, 30 kV vs 200 kV, more diffraction contrast can be seen in the LAFSD image compared with the bright field (BF) STEM image, Fig. 9 (a and b), but the particle morphology and interface structure in the LAFSD image, Fig. 9(a), is comparable with that shown in the BF STEM image. The BF TEM image is known to be sensitive to strain fields [25] induced by features like dislocations [26], surface oxides [27] and bend contours, which can make it more difficult to distinguish nanoparticles from these features or contamination spots in BF TEM image than in FSD or STEM images. In the dark field mode (DF), both the HAFSD and HAADF STEM images show obvious Z-contrast. In comparison with the HAADF STEM image formed exclusively by Z contrast, some surface features can also be seen in the HAFSD image, probably because the collection angle for the HAFSD image is not high enough and some contribution from coherent elastic scattering (diffraction) are also collected by the detectors [20, 24]. Values of imaging parameters and SNRs of these images are also summarized in Fig. 9. The SNR is calculated according to SNR=$\frac{|\bar{I}_P-\bar{I}_M|}{\sigma_M}$, where $\bar{I}_P$ is the average signal intensity of the particle, $\bar{I}_M$ and $\sigma_M$ are the average and standard deviation of signal intensity of the matrix. A threshold SNR >1 has been suggested by Österreichere et al [9] for a particle to be visible in an electron micrograph. Although the probe current and dwell time used in FSD and STEM imaging are different, for the particle of similar size, comparable values of SNR, ~10, can be seen from these images, suggesting the FSD image quality is comparable with STEM and TEM images.

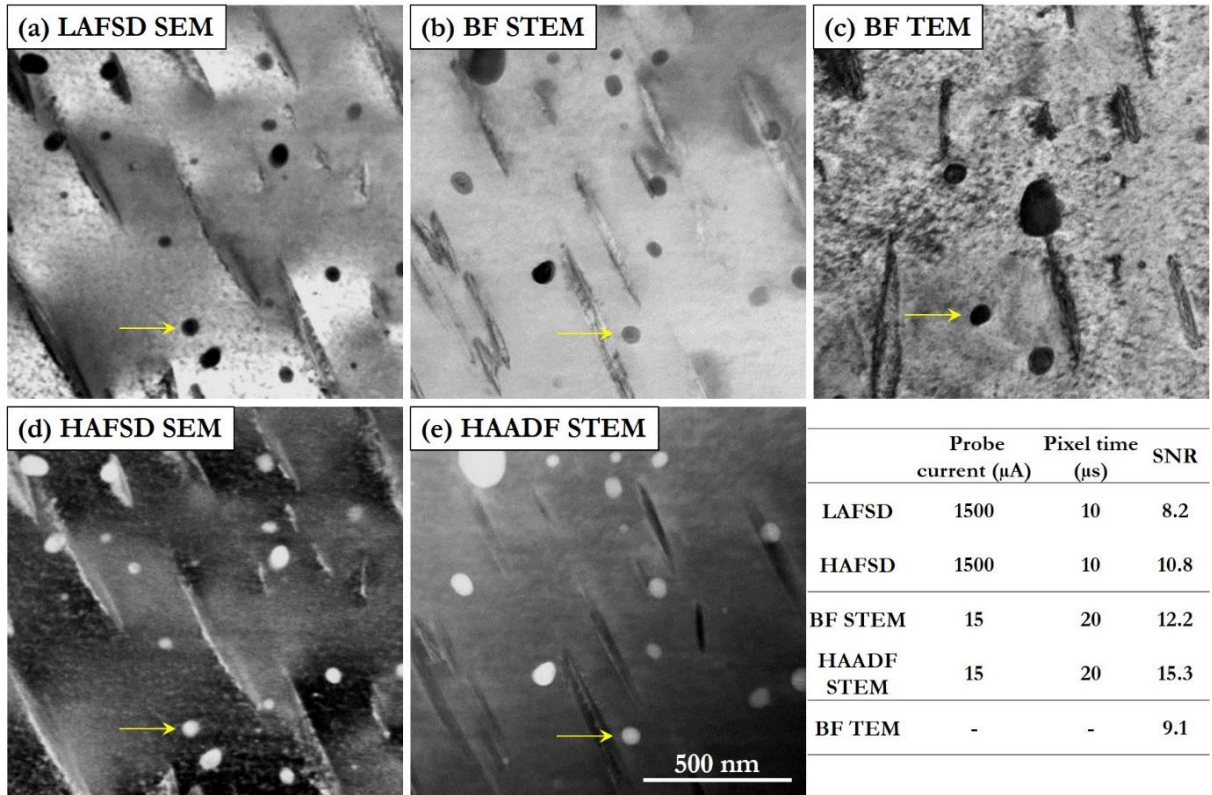

Fig. 9 Comparisons of embedded Nb particles in the Zr-Nb sample imaged by (a, d) FSD in an SEM at 30 kV, $d_d$=20 mm, (b, e) BF and HAADF STEM in a STEM at 200 kV and (c) BF TEM at 200 kV, SNRs of each image are measured and calculated based on the brightness from particles highlighted by yellow arrows and its nearby regions

Apart from characterizing embedded nanoparticles, the FSD imaging (ARGUS™) system in an SEM can also be applied to reveal other features at the nanoscale, like dislocations. The following example illustrates the application of FSD imaging to the characterisation of a similar Zr-2.5Nb alloy, Fig. 10. Determination of grain morphology and the size and distribution of second phase particles is possible by LAFSD imaging, but even smaller features such as dislocations can also be easily seen. As an illustration, in the region marked by the dashed rectangle in Fig. 10, dislocations pinning by nanoscale obstacles is clearly observed.

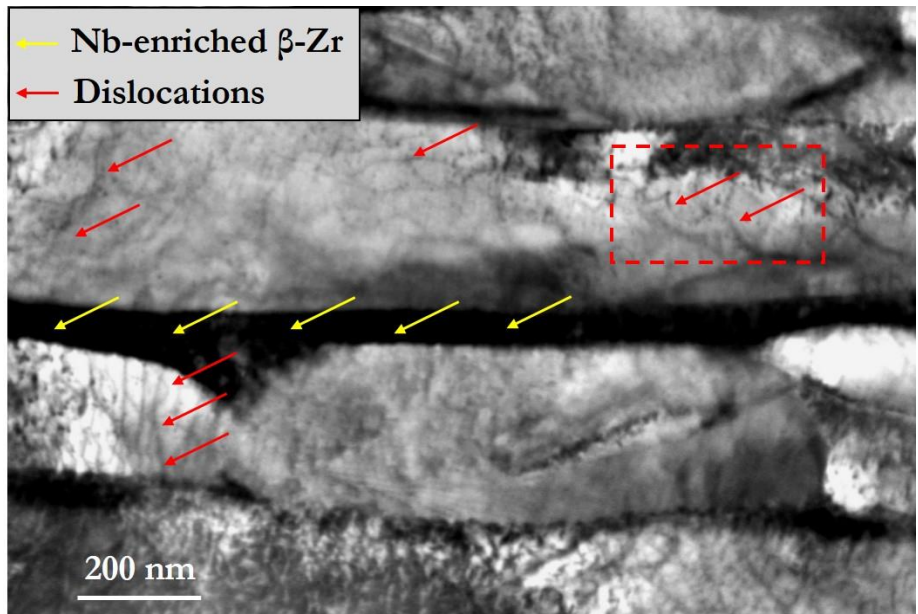

Fig. 10 An example of LAFSD ($E_0$=30 keV, $d_d$=20 mm) image showing the microstructure of a Zr-2.5Nb sample. Features such as dislocations and the second phase are highlighted by arrows.

## 4. Conclusions

These results indicate that imaging a thin foil in transmission with forescattered electrons in an SEM is suitable to investigate the microstructure of engineering materials, including the size and distribution of embedded small nanoparticles that we would normally only expect to be able to study in a TEM. By selecting suitable imaging parameters, either Z-contrast or diffraction contrast imaging mode can be selected, and the FSD image quality is comparable with STEM images but without the need for expensive instrumentation. Small features <1 nm, e.g. dislocations, can be detected reliably. Monte Carlo simulations were used to determine the critical detector distances for optimizing contrast conditions for different accelerating voltages, offering users a guideline of how to select suitable imaging parameters that can be easily applied to a much wider range of materials systems. In addition, compared with TEM and STEM, normally at high accelerating voltages, e.g. 200 and 300 kV, FSD imaging in an SEM uses much lower voltages (e.g. <10 kV) which means it is also of the capabilities to study electron beam–sensitive materials at the nano-scale.

**Acknowledgements**

The authors acknowledge the MUZIC project for providing zirconium samples. EPSRC grants (EP/K040375/1 and EP/N010868/1) are acknowledged for funding the 'South of England Analytical Electron Microscope' and the Zeiss Crossbeam FIB/SEM used in this research. We would also like to thank Dr. Daniel Goran, Bruker Nano GmbH, Germany, for helpful discussions.